\documentclass[journal]{IEEEtran}

\ifCLASSOPTIONcompsoc
 \usepackage[caption=false,font=normalsize,labelfont=sf,textfont=sf]{subfig}
\else
 \usepackage[caption=false,font=footnotesize]{subfig}
\fi

\usepackage{cite}
\usepackage{hyperref}
\usepackage{amsmath,amssymb,amsfonts}
\usepackage{algorithm}          
\usepackage{algpseudocode}
\usepackage{graphicx}
\usepackage{textcomp}
\usepackage{caption}
\usepackage{xcolor}
\usepackage{graphicx}
\usepackage{diagbox} 
\usepackage{array}   
\usepackage{amsmath, amssymb}
\usepackage{tikz}
\usetikzlibrary{arrows.meta, positioning, calc}
\usepackage{xcolor}
\usepackage{enumitem}
\usepackage{makecell}

\begin{document}

\title{A Scalable Cloud--Edge Collaborative \\CKM Construction Framework\\
Enabled by a Foundation Prior Model}

\author{Sixu~Xiao,
        Yong~Zeng,~\IEEEmembership{Fellow,~IEEE,}
        Haotian~Rong,
        and~Yanqun Tang
\thanks{S. Xiao, H. Rong, and Y. Zeng are with the National Mobile Communications Research Laboratory, Southeast University, Nanjing 210096, China (e-mail:\{230248845, 220241249, yong\_zeng\}@seu.edu.cn). Y. Zeng is also with Purple Mountain Laboratories, Nanjing 211111, China.}
\thanks{(\textit{Corresponding author: Yong Zeng.})}
\thanks{Y. Tang is with School of Electronics and Communication Engineering, Sun Yat-Sen University, Guangzhou 510006, China (e-mail:tangyq8@mail.sysu.edu.cn).}
}
\markboth{}%
{Shell \MakeLowercase{\textit{et al.}}: Bare Demo of IEEEtran.cls for IEEE Journals}
%



\maketitle

\begin{abstract}
Channel knowledge maps (CKMs) provide a site-specific, location-indexed knowledge base that supports environment-aware communications and sensing in 6G networks. In practical deployments, CKM observations are often noisy and irregular due to coverage-induced sparsity and hardware-induced linear/nonlinear degradations. Conventional end-to-end algorithms couple CKM prior information with task- and device-specific observations, and require labeled data and separate training for each construction configuration, which is expensive and therefore incompatible with scalable edge deployments. Motivated by the trends toward cloud--edge collaboration and the Artificial Intelligence--Radio Access Network (AI--RAN) paradigm, we develop a cloud--edge collaborative framework for scalable CKM construction, which enables knowledge sharing across tasks, devices, and regions by explicitly decoupling a generalizable CKM prior from the information provided by local observations. A foundation model is trained once in the cloud using unlabeled data to learn a generalizable CKM prior. During inference, edge nodes combine the shared prior with local observations. Experiments on the CKMImageNet dataset show that the proposed method achieves competitive construction accuracy while substantially reducing training cost and data requirements, mitigating negative transfer, and offering clear advantages in generalization and deployment scalability.
\end{abstract}


%
\IEEEpeerreviewmaketitle

\section{Introduction}\label{sec:Introduction}

\IEEEPARstart{S}{ix}-generation (6G) mobile communication networks are expected to shift from conventional environment-unaware architectures to environment-aware and knowledge-centric infrastructures, where network decisions are guided by previously learned knowledge of the radio environment rather than instantaneous channel state information (CSI) alone. This shift is further motivated by massive communication~\cite{itu_pr_imt2030_2023}, where a large number of intermittently active and heterogeneous devices make frequent CSI acquisition and centralized coordination increasingly costly. In large-scale deployments, such knowledge must be made available not only at centralized cloud controllers but also at distributed edge nodes, so that base stations (BSs), access points, and user equipment (UEs) can perform local and low-latency decision-making under constraints on communication, computation, and storage.

Channel knowledge maps (CKMs)~\cite{zeng2021toward} provide such a site-specific, location-indexed knowledge base. A CKM depicts the long-term structure of the channel (e.g., large-scale gain, shadowing, line-of-sight (LoS) state) and the environment (e.g., buildings and scatterers) in a spatially indexed representation. Queried at runtime, CKMs may support predictive beam selection, resource allocation, user association, and handover with reduced online overhead~\cite{zeng2024tutorial}. For a future cloud--edge collaborative 6G architecture, CKMs naturally play the role of a shared medium of radio knowledge: cloud servers can learn rich priors over CKMs from massive, diverse datasets aggregated from different regions, while distributed edge nodes exploit and refine such priors using their local observations.

Accurate and efficient CKM construction, however, remains a central challenge. In practice, observations are often noisy and irregular. This irregularity arises from heterogeneous observation degradations: (i) \emph{coverage-induced} degradations due to incomplete spatial sampling, leading to missing entries or effective low-resolution observations; and (ii) \emph{hardware-induced} degradations caused by sensing front-end limitations, such as signal truncation and quantization. These factors make CKM construction an ill-posed inverse problem~\cite{daras2024survey}.
 Moreover, observations are generated and consumed in a spatially distributed manner by many BSs and sensing nodes that may operate under hardware constraints and limited backhaul capacities, which forces each BS to handle multiple CKM construction tasks. A purely centralized solution would incur prohibitive latency and backhaul overhead in large networks, while a straightforward fully distributed deployment, where each BS maintains full models for different CKM construction scenarios, leads to dramatically increased computing and memory cost as the number of BSs grows.

In the remainder of Section~\ref{sec:Introduction}, we review existing CKM construction methods and motivate the proposed paradigm. Section~\ref{sec:system model} introduces representative CKM construction scenarios and methods. Section~\ref{sec:The_Proposed_Framework_and_Method} presents the proposed framework and realizes it via a foundation score-based model together with plug-and-play inference. Section~\ref{sec:comparative_study} reports a comparative study and experimental results. Section~\ref{sec:conclusion} concludes the paper.

\subsection{Related Works and Gaps}
Several existing methods attempt to solve the inverse problem of CKM construction using
end-to-end neural networks, such as U-Net~\cite{levie2021radiounet} and Transformer-based models~\cite{tian2021transformer}, which map the physical environment and BS locations to CKMs in a supervised manner. While such models can handle specific tasks with high accuracy, they cannot generalize across tasks and devices, i.e., the model must be retrained or fine-tuned for each new task or hardware configuration, which is too costly for scalable edge deployments.

Recent advances in generative artificial intelligence (GAI) have introduced diffusion- and score-based models into CKM construction by sampling from the posterior distribution conditioned on observations. For example, \cite{fu2025ckmdiff} proposes the CKMDiff model, which employs a
conditional decoupled diffusion model \cite{huang2023decoupled} to construct the complete CKM using partially
observed channel knowledge data. It directly models the posterior distribution and samples from it; however, this approach faces several
challenges. First, it models the posterior distribution via a black-box algorithm and lacks generalization across tasks and devices, similar to supervised methods. Second, sampling-based generative methods have a trade-off between diversity
and fidelity~\cite{dhariwal2021diffusion}, and the desired behavior fundamentally depends on the type of tasks. 
Many natural image restoration problems are \emph{one-to-many}: the observation can correspond to multiple perceptually plausible reconstructions, and diversity is often acceptable or even encouraged~\cite{zheng2019pluralistic}. In contrast, a CKM construction inverse problem is \emph{correctness-oriented}:   it prefers a single correct construction solution, and diversity is often undesirable. \cite{wang2025radiodiff} integrates Bayesian
inverse estimation based on a diffusion model pretrained on natural images to
perform maximum a posteriori (MAP)-based filtering for constructing the complete
CKM from sparse and noisy data. Although the method in \cite{wang2025radiodiff}
requires no training on CKM data, this paradigm can only
exploit generic low-level spatial priors, lacking the capacity
to integrate domain-specific knowledge, and also remains
restricted to single-task settings.

To date, only limited attention has been paid to cloud--edge collaborative CKM construction. CollaboRadio~\cite{shao2025collaboradio} proposes a hybrid device-edge-cloud paradigm for fine-grained CKM construction under ultra-sparse sampling. It constructs local CKM patches at the edge with lightweight models, and fuses these patches in the cloud to form a global map, thereby offloading computation and reducing uplink traffic. However, its cloud--edge collaboration is mainly considered from the perspective of data acquisition and aggregation, rather than a system-level division of labor between training and inference. For instance, the edge is naturally suited for inference due to its proximity to where measurements are generated. In contrast, the cloud with substantially stronger computing resources is better positioned for training. Moreover, CollaboRadio remains an end-to-end pipeline whose edge and cloud models are trained on a configuration-specific dataset in a supervised manner, which cannot adapt to heterogeneous observation degradations. As a result, scaling to new configurations typically requires a new labeled dataset and retraining the pipeline, limiting the scalability of deployment.

Overall, existing approaches suffer from several limitations: (i) high training cost due to separate model training for each construction configuration; (ii) the need for large amounts of labeled training data for each task and sensing device configuration; (iii) limited generalization across different network configurations; (iv) centralized training and inference that do not fully exploit task- and device-agnostic knowledge sharing. More fundamentally, most existing CKM construction studies are \emph{algorithm-centric}: they primarily optimize construction accuracy for a given task and a fixed sensing configuration, while leaving it unexplored how CKM construction should be organized and scaled at a \emph{system level}.  As the AI--RAN~\cite{khan2023ai} paradigm evolves, network services are moving toward cloud-native operation with on-demand scaling, and edge nodes (e.g., BSs) are expected to have non-negligible computing capabilities. It is therefore important to make CKM construction \textbf{scalable} and leverage edge computing resources.

\subsection{Proposed Approach Overview and Contributions}
From a  system-level viewpoint, two observations are crucial. First, the spatial–spectral regularities of the channel embedded in CKMs are transferable across different CKM construction tasks and regions, and are agnostic to the sensing devices. Second, such a prior can be learned once in the cloud and then shared with many edge nodes within a cloud--edge architecture, and this prior needs to be incorporated with local observations only at inference time, which suggests that instead of training separate end-to-end models, it is more desirable to (i) explicitly decouple the modeling of the CKM prior knowledge from the information provided by specific local observations, and (ii) deploy the learned prior as a \textbf{reusable and centralized} knowledge base that can be distributed to network edges for inference in a plug-and-play manner.

\begin{figure}
    \centering
    \includegraphics[width=0.9\linewidth]{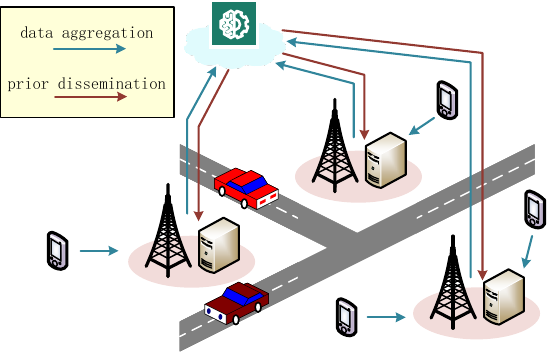}
    \caption{An illustration of cloud--edge knowledge sharing for CKM construction.}
    \label{fig:paradigm_small}
\end{figure}

A foundation model~\cite{bommasani2021opportunities} serves as a natural choice for realizing such a reusable prior, which is learned and maintained in the cloud and queried at the edges. As illustrated in Fig.~\ref{fig:paradigm_small}, we develop a cloud--edge collaborative framework for scalable CKM construction that explicitly enables knowledge sharing. We instantiate this framework with a CKM construction algorithm based on a foundation score-based model, which can learn the complex distribution of a high-dimensional vector. A score-based model is first trained on an unlabeled CKM dataset to learn the observation-agnostic prior knowledge. In the inference stage, edge nodes perform posterior inference that combines the shared prior with their locally measured, irregular observations. This leads to a scalable CKM construction paradigm that explicitly enables knowledge sharing across tasks, regions, and device front ends.

The main contributions are summarized as follows.
\begin{itemize}
  \item \textbf{Scalable cloud--edge collaborative CKM construction framework:} We propose a system-level cloud--edge collaborative framework for scalable CKM construction, where a generalizable CKM prior is learned once, maintained in the cloud, and reused by multiple edge nodes. We further show that generative AI is essential in this framework because it enables explicit decoupling between the modeling of shared prior knowledge and the information provided by local observations, and it is able to capture the complex distribution of high-dimensional CKMs. This formulation clarifies the respective roles of the cloud (data aggregation and prior learning) and the edge (task- and device-specific inference) and highlights the benefits of knowledge sharing in terms of training cost, data requirements, and deployment scalability.
  \item \textbf{Foundation score-based prior model and plug-and-play posterior inference algorithm:} We instantiate
    this framework with a score-based foundation model that learns the observation-agnostic CKM prior from unlabeled data in the cloud and decouples this prior from task- and device-specific observation likelihoods. Building upon diffusion posterior sampling (DPS)~\cite{chung2023diffusion}, we utilize a plug-and-play reverse stochastic differential equation (SDE) that combines the learned prior with any differentiable degradation operators, enabling edge nodes to perform MAP-oriented posterior inference for various CKM construction configurations with zero-shot adaptation.
  \item \textbf{Comprehensive and controlled evaluations:} Using the CKMImageNet dataset~\cite{wu2025ckmimagenet}, we conduct extensive experiments across multiple CKM construction configurations and analyze the proposed approach in terms of efficiency and accuracy. The results show that the proposed method achieves competitive construction accuracy with high scalability and significantly reduces training cost and data requirements compared to end-to-end baselines. In addition, we analyze the sensitivity of the proposed algorithm to the key hyperparameter and provide a guideline for its selection. The source code is available at \href{https://github.com/xxxxssxx/Cloud-Edge-Collaborative-CKM-Construction}{https://github.com/xxxxssxx/Cloud-Edge-Collaborative-CKM-Construction}.
\end{itemize}

\begin{figure*}
    \centering
    \includegraphics[width=0.9\linewidth]{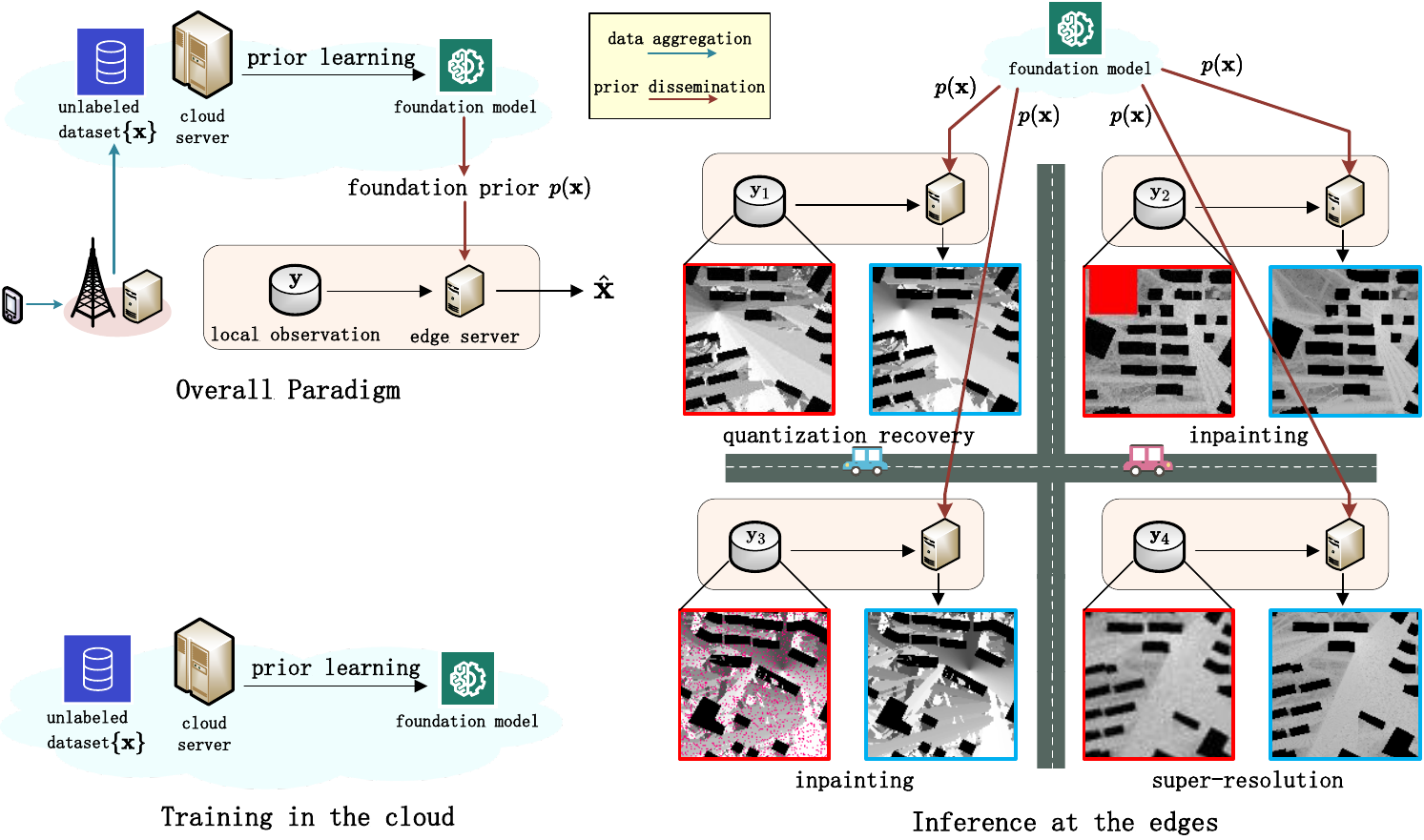}
    \caption{Proposed cloud--edge collaborative knowledge-sharing CKM construction framework.}
    \label{fig:paradigm}
\end{figure*}

\section{System Model and Problem Formulation}\label{sec:system model}

In this paper, we consider the construction of a CKM for a given area from noisy and irregular channel knowledge observations. The area is partitioned into patches, each with $n$ grid cells. The CKM is therefore represented by a vector $\mathbf{x} \in \mathbb{R}^{n}$. Let $\mathbf{y}\in \mathbb{R}^{n'}$ denote the channel knowledge observation and $\mathcal{A}(\cdot)$ denote a (possibly nonlinear) forward operator that captures the observation process. The measurement noise is modeled as additive white Gaussian noise $\mathbf{n} \sim \mathcal{N}(\mathbf{0}, \sigma^2 \mathbf{I})$. The observation model is thus
\begin{equation}\label{eq:observation_process}
    \mathbf{y} = \mathcal{A}(\mathbf{x}) + \mathbf{n}.
\end{equation}
Recovering $\mathbf{x}$ from $\mathbf{y}$ under Eq.~\eqref{eq:observation_process} forms an ill-posed inverse problem. 
\subsubsection{Representative Observation Operators}
The forward operator $\mathcal{A}(\cdot)$ can model both coverage-induced and hardware-induced degradations in CKM construction problems. 

\paragraph{Masking}\label{paragraph:masking}
 When the CKM data are missing at some grid cells, the operator $\mathcal{A}(\cdot)$ is a binary diagonal matrix $\mathbf{A}\in\{0,1\}^{n\times n}$, whose diagonal entry $A_{ii}=1$ indicates that the $i$-th element of $\mathbf{x}$ is observed and $A_{ii}=0$ otherwise. The corresponding CKM construction task is inpainting.

\paragraph{Downsampling}\label{paragraph:downsampling}
When the CKM data are collected with coarse spatial sampling (e.g., sparse user trajectories or low-density sensing), the operator $\mathcal{A}(\cdot)\triangleq\mathbf{D}_{s}(\cdot)$ denotes a linear downsampling operator with a known scale factor $s>1$. The corresponding CKM construction task is super-resolution. 

\paragraph{Dynamic-range truncation}\label{paragraph:truncation}
To model the saturation or limited dynamic range of sensing devices, $\mathcal{A}(\cdot)$ can be an element-wise truncation operator
\begin{equation}\label{eq:truncation_operator}
    \bigl[\mathcal{A}_{\mathrm{tr}}(\mathbf{x})\bigr]_i
    =
    \min\bigl(\max(x_i, a),\, b\bigr),
\end{equation}
where $a$ and $b$ denote the lower and upper truncation thresholds, respectively. The corresponding CKM construction task is truncation recovery. 

\paragraph{Finite-resolution quantization}\label{paragraph:quantization}
To model the finite measurement resolution of sensing devices, $\mathcal{A}(\cdot)$ can be an element-wise quantization operator
\begin{equation}
    \bigl[\mathcal{A}_{\mathrm{q}}(\mathbf{x})\bigr]_i
    =
    Q(x_i),
\end{equation}
where $Q(\cdot)$ maps a continuous value to a discrete reproduction level according to a set of quantization bins. The corresponding CKM construction task is quantization recovery. 

Overall, masking and downsampling are linear degradation operators, while truncation and quantization are nonlinear degradation operators.

\subsubsection{MMSE- and MAP-Based CKM Construction}

From a Bayesian viewpoint, let $p(\mathbf{x})$ denote the prior distribution of the CKM and $p(\mathbf{y}\vert\mathbf{x})$ denote the likelihood of obtaining an observation $\mathbf{y}$ given the underlying CKM $\mathbf{x}$. The posterior distribution is
\begin{equation}\label{eq:Bayesian}
    p(\mathbf{x}\vert\mathbf{y}) \propto p(\mathbf{x})\,p(\mathbf{y}\vert\mathbf{x}).
\end{equation}
Depending on how this posterior is exploited, two standard construction criteria are commonly used.

\paragraph{MMSE-style construction}\label{para:MMSE_construction}
A widely adopted approach is to learn a mapping $f(\cdot)$ from observation $\mathbf{y}$ to CKM $\mathbf{x}$ and minimize a suitable expected loss between the estimated CKM $\hat{\mathbf{x}}$ and the ground truth $\mathbf{x}$. Let $\mathcal{L}(\cdot,\cdot)$ denote a discrepancy between the two CKMs. The MMSE-style construction problem can be written as
\begin{equation}
  \begin{aligned}
    \min_{f}\quad & \mathbb{E}\big[\mathcal{L}(\hat{\mathbf{x}}, \mathbf{x})\big] \\[0.5ex]
    \text{s.t.}\quad
    & \hat{\mathbf{x}} = f(\mathbf{y}),\\
    & \mathbf{y} = \mathcal{A}(\mathbf{x}) + \mathbf{n},
  \end{aligned}
\end{equation}
where the expectation is taken over the joint distribution of $(\mathbf{x},\mathbf{y})$ and $\mathcal{L}$ is the mean-square error (MSE). The optimal estimator is $\hat{\mathbf{x}}(\mathbf{y}) = \mathbb{E}\left[\mathbf{x}\vert\mathbf{y}\right]=\int\mathbf{x}\,p\left(\mathbf{x}\left\vert\mathbf{y}\right.\right)d\mathbf{x}$~\cite{bishop2006prml}. 

\paragraph{MAP-style construction}
Another principled approach is the MAP estimation, which searches for the most probable CKM given the observation $\mathbf{y}$. Under the Gaussian noise model in Eq.~\eqref{eq:observation_process}, MAP construction solves
\begin{equation}
    \begin{split}
          \hat{\mathbf{x}}_{\mathrm{MAP}}(\mathbf{y})
  &= \arg\max_{\mathbf{x}} \; p(\mathbf{x}\vert\mathbf{y}) \\
  &= \arg\max_{\mathbf{x}} \Big\{ \log p(\mathbf{x}) + \log p(\mathbf{y}\vert\mathbf{x}) \Big\}, \\
  &= \arg\min_{\mathbf{x}} \left\{
    -\log p(\mathbf{x}) + \frac{1}{2\sigma^2}\big\|\mathbf{y}-\mathcal{A}(\mathbf{x})\big\|_2^2
  \right\}.
    \end{split}
\end{equation}

\section{The Proposed Framework and Method}\label{sec:The_Proposed_Framework_and_Method}

\subsection{Problem Analysis}\label{subsec:problem_analysis}

In practical 6G deployments, channel knowledge observations are predominantly obtained at the network edge by BSs, access points, and UEs. Thus, a BS is required to handle multiple CKM construction tasks simultaneously, under heterogeneous sensing hardware and deployment conditions. Uploading massive raw measurements to the cloud for every construction task instance is impractical due to backhaul limitations, and may also raise privacy and security concerns. Meanwhile, cloud infrastructures can provide abundant computing and storage resources for training large models. This naturally motivates a cloud--edge collaborative CKM construction framework: train powerful models in the cloud, and perform inference close to where data are generated~\cite{zhou2019edge}.  A central design objective for achieving deployment scalability under this architecture is \emph{knowledge sharing}: instead of training a separate model for every task, every device, and every noise level, we would like to identify and reuse the common knowledge that is invariant across tasks and devices. 

As discussed in Section~\ref{sec:system model}, the posterior distribution governs both MMSE-style construction and MAP-style construction. As shown in Eq.~\eqref{eq:Bayesian}, the posterior distribution $p(\mathbf{x}\vert\mathbf{y})$ is proportional to $p(\mathbf{x})p(\mathbf{y}\vert\mathbf{x})$, where $p(\mathbf{x})$ is the observation-agnostic CKM prior and $p(\mathbf{y}\vert\mathbf{x})$ is the task- and device-specific likelihood. This factorization clarifies \textbf{what can be shared}: the prior $p(\mathbf{x})$ captures the stable, generalizable features of the radio environment, such as building layouts and propagation statistics across different regions; while the likelihood $p(\mathbf{y}\vert\mathbf{x})$ encodes local sensing geometry and hardware characteristics, and thus varies across tasks and devices. Consequently, any form of knowledge sharing across tasks should be realized at the level of the prior $p(\mathbf{x})$; the likelihood is inherently specific to each inverse problem instance.

This observation also clarifies the limitations of existing
paradigms. Classical interpolation-based methods implicitly assume trivial priors (e.g., local smoothness), usually leading to poor construction quality, and are hard to use in nonlinear construction tasks where amplitude information is destroyed. Moreover, they rely solely on the instantaneous observation $\mathbf{y}$ and therefore cannot support knowledge sharing. MMSE-style construction improves upon this by learning a deterministic mapping $f:\mathbf{y}\mapsto\hat{\mathbf{x}}$ that minimizes an expected loss. In practice, this can be implemented by end-to-end supervised models trained on labeled data. Likewise, conditional generative approaches, which directly model the entire posterior $p(\mathbf{x}\vert\mathbf{y})$ and sample from it, also learn a specialized conditional model.  However, such models couple the prior $p(\mathbf{x})$ and the likelihood $p(\mathbf{y}\vert\mathbf{x})$ inside a single black-box mapping. When the task, degradation operator, or sensing configuration changes, the entire model typically needs to be retrained or fine-tuned with new labeled data, and the reusable part of learned knowledge is not explicitly extracted or shared.

In contrast, a MAP-style formulation naturally supports knowledge sharing. By keeping $p(\mathbf{x})$ and $p(\mathbf{y}\vert\mathbf{x})$ as separate modeling components, one can first learn a high-capacity prior $p(\mathbf{x})$ from a large unlabeled CKM dataset and then combine it with different likelihoods at inference time via
\begin{equation}
    \hat{\mathbf{x}}_{\mathrm{MAP}}(\mathbf{y})
    = \arg\max_{\mathbf{x}} \big\{ \log p(\mathbf{x}) + \log p(\mathbf{y}\vert\mathbf{x}) \big\}.
\end{equation}
Therefore, if the objective is to realize cross-task, cross-device CKM knowledge sharing and scalable CKM construction, it is essential to explicitly model the prior distribution $p(\mathbf{x})$ instead of hiding it inside configuration-specific end-to-end models.

\subsection{Scalable Cloud--Edge Collaborative CKM Construction Framework via Generative AI}
\label{subsec:framework}

Building on the above analysis, we propose a scalable cloud--edge collaborative knowledge-sharing CKM construction framework enabled by a foundation prior model, as shown in Fig.~\ref{fig:paradigm}.
\paragraph*{Roles of cloud and edge}
The cloud is responsible for learning, maintaining, and periodically refreshing a CKM prior $p(\mathbf{x})$ from large-scale aggregated unlabeled data, leveraging its abundant computing and storage resources.  Edge nodes (BSs equipped with edge servers) are responsible for lightweight inference by defining the task- and device-specific likelihood $p(\mathbf{y}\vert\mathbf{x})$ and incorporating it with the prior $p(\mathbf{x})$. Because observations are generated at BSs/UEs and may be costly or privacy-sensitive to transmit over the backhaul, the posterior inference is performed at the network edge, as close to the data source as possible. This cloud--edge separation turns the CKM prior into a reusable knowledge base that can be shared across tasks, devices, and regions, while keeping configuration instantiation and inference local and scalable.

\paragraph*{Workflow}
The resulting workflow follows the principle of ``train once in the cloud, adapt everywhere at the edge'':
\begin{enumerate}
    \item \textbf{Unlabeled data aggregation:} CKM data collected across different regions and networks are aggregated in the cloud to form a large unlabeled dataset $\{\mathbf{x}\}$. The aggregated data may contain heterogeneous quality levels, including clean CKM snapshots as well as imperfect CKM samples.
    \item \textbf{Cloud pretraining and refreshing:}  The cloud trains a generative foundation model to learn the CKM prior $p(\mathbf{x})$ from the large unlabeled dataset $\{\mathbf{x}\}$~\cite{ho2020denoising, lipman2023flow,song2021score,daras2023ambient}. As additional data accumulate over time (e.g., from new regions, seasons, or network upgrades), the cloud performs continual pretraining or periodic refreshing to improve the prior and track long-term environmental changes.
    \item \textbf{Edge posterior inference:} Given a local observation $\mathbf{y}$ under coverage limitations and/or hardware constraints, the edge node specifies the corresponding forward operator $\mathcal{A}(\cdot)$ and likelihood  $p(\mathbf{y}\vert\mathbf{x})$~\cite{cranmer2020frontier,chung2023diffusion,song2023pseudoinverse}. It then performs posterior inference to obtain the constructed CKM by combining the shared prior $p(\mathbf{x})$ with the local likelihood $p(\mathbf{y}|\mathbf{x})$ in a plug-and-play manner~\cite{wang2023zeroshot,venkatakrishnan2013plug}, so that changing tasks or device front ends affects $\mathcal{A}(\cdot)$ and the likelihood term, rather than the shared prior.
    \item \textbf{Model distribution and evolution:} Updated foundation model weights (or their compressed/quantized versions) are disseminated from the cloud to edge nodes and cached at edge servers when appropriate~\cite{huang2024mobile,khan2023ai}. This enables scalable deployment, versioned model evolution, and long-term compatibility with newly introduced tasks and heterogeneous sensing hardware.
\end{enumerate}

\paragraph*{Why generative AI}

Generative modeling provides an explicit, reusable
representation of $p(\mathbf{x})$ that can be combined with arbitrary likelihoods at inference time, which enables scalable CKM construction. Downstream construction tasks are then treated as Bayesian inverse problem instances, where the shared prior is combined with task- and device-specific observation processes.

\subsection{Cloud-side Foundation Score Model and Training Method}\label{section:cloud_side}
We implement the cloud-side foundation prior model using a score-based generative model~\cite{song2021score}. In classical wireless inverse problems (e.g., channel estimation), priors are typically imposed through tractable statistical or structural assumptions, such as Gaussian/Rician models with channel covariance information, or sparsity in the angle-delay domain. However, CKM construction deals with a high-dimensional spatial field with environment-dependent and non-stationary structures, where such hand-crafted priors become hard to generalize.
  The score-based model, which is a more general concept than the diffusion model~\cite{ho2020denoising}, can be used to model a complex distribution of a high-dimensional variable and sample from it. An SDE can smoothly transform a data distribution $p(\mathbf{x})$ (i.e., the distribution of CKM) into a known distribution (i.e., the Gaussian distribution) by slowly injecting noise, and a corresponding reverse-time SDE can transform the known distribution back into the data distribution by slowly removing the noise. The forward SDE and corresponding reverse SDE~\cite{anderson1982reverse} take the form of
\begin{subequations}\label{SDE}
\begin{align}
    \mathrm{d}\mathbf{x} &= \mathbf{f}(\mathbf{x}, t) \mathrm{d}t + g(t)\mathrm{d}\mathbf{w} \label{SDE:forward} \\
    \mathrm{d}\mathbf{x} &= \left[ \mathbf{f}(\mathbf{x}, t) - g^2(t)\nabla_{\mathbf{x}}\log p(\mathbf{x})\right]\mathrm{d}t + g(t)\mathrm{d}\mathbf{w}, \label{SDE:reverse}
\end{align}
\end{subequations}
where $\mathbf{f}(\mathbf{x}, t)$ is called the drift coefficient, $g(t)$ is a time-dependent diffusion coefficient, $\mathrm{d}\mathbf{w}$ represents a standard Wiener process, $t \in \left[0, T\right]$ is a continuous time variable where $t=0$ is the start of the corruption process, and therefore $\mathbf{x}_{0}$ stands for the ground truth. The score $\nabla_{\mathbf{x}_t} \log p(\mathbf{x}_t)$ indicates the direction in which the latent CKM sample $\mathbf{x}_t$ should be updated to move toward regions of higher probability density under $p(\mathbf{x}_t)$.

\begin{figure}
    \centering
    \includegraphics[width=0.9\linewidth]{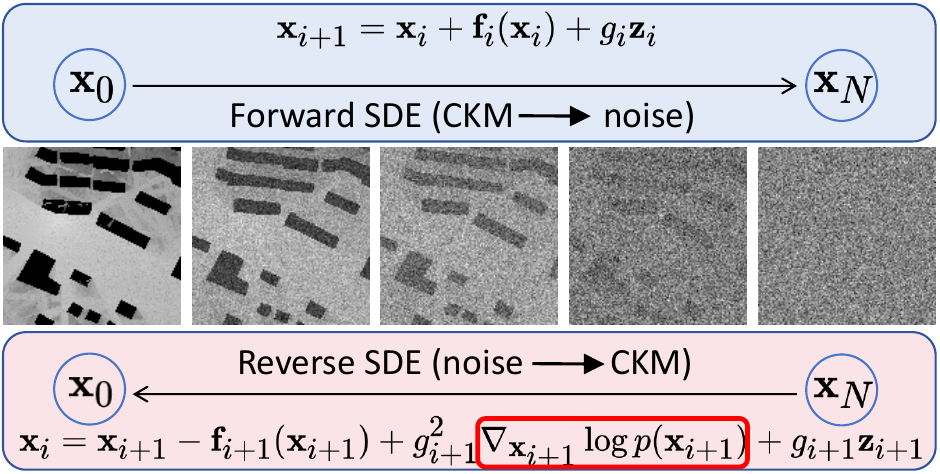}
    \caption{Score-based CKM construction, by solving the reverse SDE to draw a sample from $p(\mathbf{x})$ after learning the score $\nabla_{\mathbf{x}_{i}}\log p(\mathbf{x}_{i})$ at every timestep $i$.
}
    \label{fig:algorithm_visualization_training}
\end{figure}

An SDE can be reversed if the score $\nabla_{\mathbf{x}_{t}} \log p(\mathbf{x}_{t})$ is known. One can obtain a sample $\mathbf{x}$ that follows the distribution $p(\mathbf{x})$ by solving the reverse SDE. A schematic diagram is shown in Fig.~\ref{fig:algorithm_visualization_training}.
The score can be approximated using the score matching \cite{song2019generative} method by training a time-dependent neural network, that is, $\mathbf{s}_{\theta}(\mathbf{x}_{t}, t) \approx \nabla_{\mathbf{x}_{t}} \log p(\mathbf{x}_{t})$. A U-Net \cite{ronneberger2015u} is chosen as the backbone. There are various types of SDE, such as the Variance Preserving (VP) SDE and the Variance Exploding (VE) SDE. For the VP SDE, the forward SDE is written as 
\begin{equation}
\mathrm{d}\mathbf{x} = -\frac{1}{2}\beta(t)\mathbf{x}\mathrm{d}t + \sqrt{\beta(t)} \mathrm{d}\mathbf{w},\quad t\in \left[0, T\right]
\end{equation}
where the corresponding drift coefficient and diffusion coefficient are $\mathbf{f}(\mathbf{x}, t) = -\frac{1}{2}\beta(t)\mathbf{x}$ and $g(t)=\sqrt{\beta(t)}$. By discretizing it into $N$ steps, the discretized SDE defines a Markov process, whose transition kernel is given by
\begin{equation}
\mathbf{x}_{i} = \sqrt{1-\beta_{i}}\mathbf{x}_{i-1} + \sqrt{\beta_{i}}\mathbf{z}_{i-1},\quad i=1,\dots,N
\end{equation} 
where $\mathbf{z}_{i-1}\sim \mathcal{N}(0, \mathbf{I})$. The corresponding perturbation kernel can be derived as 
\begin{equation}
            \mathbf{x}_{i} = \sqrt{\bar{\alpha}_{i}}\mathbf{x}_{0} + \sqrt{1-\bar{\alpha}_{i}}\mathbf{z}_{0},\quad i=1,\dots,N
\end{equation}
where $\alpha_{i}=1-\beta_{i}$ and $\bar{\alpha}_{i} = \prod_{j=1}^{i}\alpha_{i} $. In fact, the DDPM\cite{ho2020denoising} is the discretized VP SDE case of the score-based model. According to Tweedie's formula\cite{efron2011tweedie}, the relation between the injected source noise $\mathbf{z}_0$ and the score for the VP SDE is 

\begin{equation}
    \nabla_{\mathbf{x}_{i}}\log p(\mathbf{x}_{i}) = -\frac{\mathbf{z}_0}{\sqrt{1-\bar{\alpha}_{i}}}.
\end{equation}
To model the distribution of CKM $p(\mathbf{x})$, one can train a neural network using the score matching method. The training algorithm for the VP SDE is given in Algorithm \ref{training VP}. 

\begin{algorithm}[H]
  \caption{Training the CKM foundation score model in the cloud}\label{training VP}
  \begin{algorithmic}[1]
    \Require unlabeled CKM dataset $\{\mathbf{x}\}$, noise schedule $\{\bar{\alpha}_{i}\}_{i=1}^{N}$, number of timesteps $N$.
    \Repeat
      \State $\mathbf{x}_0\sim \{\mathbf{x}\}$, $i\sim \text{Uniform}(\{1,\dots,N\})$, $\mathbf{z}_0\sim\mathcal{N}(0,\mathbf{I})$
      \State $\mathbf{x}_{i} \gets \sqrt{\bar{\alpha}_{i}}\mathbf{x}_{0} + \sqrt{1-\bar{\alpha}_{i}}\mathbf{z}_{0}$  \Comment{Perturbation Kernel}
      \State $\nabla_{\mathbf{x}_{i}}\log p(\mathbf{x}_{i}) \gets -\frac{\mathbf{z}_0}{\sqrt{1-\bar{\alpha}_{i}}}$ \Comment{Tweedie's Formula}
      \State Take gradient descent step on
      $\nabla_{\theta}\lVert\nabla_{\mathbf{x}_{i}}\log p(\mathbf{x}_{i})-\mathbf{s}_{\theta}(\mathbf{x}_{i},i)\rVert_{2}^{2}$
    \Until{converged}
  \end{algorithmic}
\end{algorithm}

\subsection{Edge-side Plug-and-play Posterior Inference Method}

\begin{figure}
    \centering
    \includegraphics[width=0.9\linewidth]{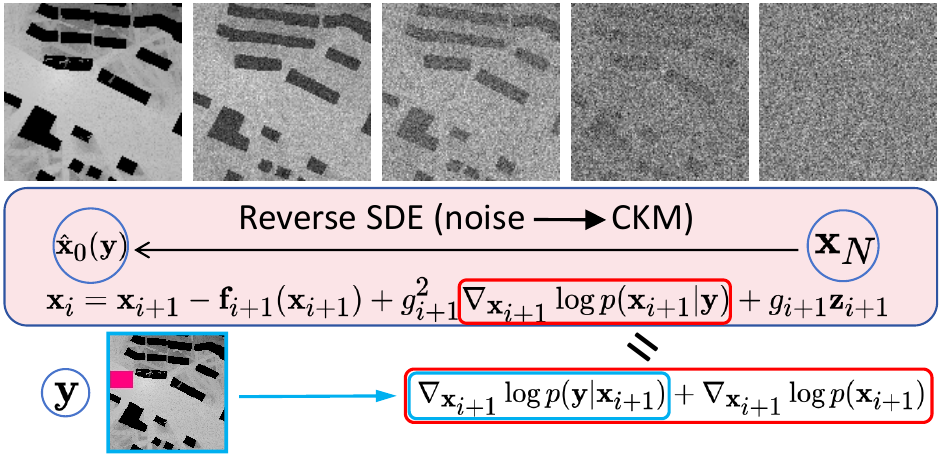}
    \caption{CKM construction guided by observation $\mathbf{y}$ that draws a sample from $p(\mathbf{x}|\mathbf{y})$. The prior score $\nabla_{\mathbf{x}_{i}}\log p(\mathbf{x}_{i})$ is replaced with the posterior score $\nabla_{\mathbf{x}_{i}}\log p(\mathbf{x}_{i}|\mathbf{y})$.
}
    \label{fig:algorithm_visualization_sampling}
\end{figure}

At the edge, we adopt diffusion posterior sampling~\cite{chung2023diffusion} to inject local observation constraints into the learned prior to perform posterior inference. The learned score function $\mathbf{s}_{\theta}(\mathbf{x}_{i},i)$ can be plugged into the reverse transition kernel, which is derived from the discretized reverse SDE in Eq.~\eqref{SDE:reverse}. Solving this Markov process is called ancestral sampling \cite{song2021score}, and it generates samples from the prior distribution $p(\mathbf{x})$. Specifically, the reverse transition kernel for the VP SDE is as follows:
\begin{equation}\label{eq:reverse_transition_kernel}
    \begin{split}
                \mathbf{x}_{i}&=\frac{1}{\sqrt{\alpha_{i+1}}} \mathbf{x}_{i+1}+\frac{1-\alpha_{i+1}}{\sqrt{\alpha_{i+1}}} \mathbf{s}_{\theta}\left(\mathbf{x}_{i+1}, i+1\right) \\
        &+\sqrt{\frac{(1-\alpha_{i})(1-\bar{\alpha}_{i-1})}{1-\bar{\alpha}_{i}}} \mathbf{z}. \quad i=N-1, \dots, 0
    \end{split}
\end{equation}
For CKM construction, the objective is to sample from the posterior distribution $p(\mathbf{x}\vert\mathbf{y})$ rather than $p(\mathbf{x})$.
This is accomplished by replacing the prior score in Eq.~\eqref{eq:reverse_transition_kernel}  with the posterior score and solving the modified Markov process. The central challenge, therefore, is to fuse the CKM prior $p(\mathbf{x})$ with various observations $\mathbf{y}$ at the edges in a plug-and-play manner. More specifically, solving the modified Markov process requires the posterior score $\nabla_{\mathbf{x}_{i}}\log p(\mathbf{x}_{i}\vert\mathbf{y})$ at every timestep $i$. A schematic diagram is shown in Fig.~\ref{fig:algorithm_visualization_sampling}.
According to Bayes' theorem, the posterior score can be expanded as
\begin{equation}
    \nabla_{\mathbf{x}_{i}}\log p(\mathbf{x}_{i}\vert \mathbf{y}) = \nabla_{\mathbf{x}_{i}}\log p(\mathbf{x}_{i}) + \nabla_{\mathbf{x}_{i}}\log p(\mathbf{y}\vert\mathbf{x}_{i} ).
\end{equation}
The prior score $\nabla_{\mathbf{x}_{i}}\log p(\mathbf{x}_{i})$ has already been learned by the foundation model $\mathbf{s}_{\theta}(\mathbf{x}_{i},i)$ in the cloud, but the likelihood score $\nabla_{\mathbf{x}_{i}}\log p(\mathbf{y}\vert\mathbf{x}_{i})$ is unknown. \cite{chung2023diffusion} proposes a computable approximation of the likelihood score. For an observation process in Eq.~\eqref{eq:observation_process} that has an arbitrary (possibly nonlinear) forward operator $\mathcal{A}(\cdot)$ and additive Gaussian noise $\mathbf{n} \sim \mathcal{N}(0, \sigma^2\mathbf{I})$, the likelihood score can be approximated as 
\begin{equation}\label{posterior_approximation}
\begin{split}
        \nabla_{\mathbf{x}_{i}}\log p(\mathbf{y}\vert\mathbf{x}_{i}) &\approx \nabla_{\mathbf{x}_{i}}\log p(\mathbf{y}\vert\hat{\mathbf{x}}_{0}) \\
        &= -\frac{1}{2\sigma^{2}} \nabla_{\mathbf{x}_{i}}\lVert \mathbf{y} - \mathcal{A}(\hat{\mathbf{x}}_{0}) \rVert_{2}^{2},
\end{split}
\end{equation}
where $\hat{\mathbf{x}}_{0}$ is the progressive estimate \cite{ho2020denoising}, which is a function of $\mathbf{x}_{i}$ and the prior score $\nabla_{\mathbf{x}_{i}}\log p(\mathbf{x}_{i})$. The only requirement on $\mathcal{A}(\cdot)$ is differentiability, so this approximation is applicable for a wide range of construction configurations, aligning with the plug-and-play, knowledge-sharing setting. Finally, the posterior score can be approximated as 
\begin{equation}
    \begin{split}
        \nabla_{\mathbf{x}_{i}}\log p(\mathbf{x}_{i}\vert\mathbf{y}) &= \nabla_{\mathbf{x}_{i}}\log p(\mathbf{x}_{i}) + \nabla_{\mathbf{x}_{i}}\log p(\mathbf{y}\vert\mathbf{x}_{i} ) \\
        &\approx \mathbf{s}_{\theta}(\mathbf{x}_{i},i) + \nabla_{\mathbf{x}_{i}}\log p(\mathbf{y}\vert\hat{\mathbf{x}}_{0}).
    \end{split}
\end{equation}
One can obtain the posterior sampling process by replacing the prior score in Eq.~\eqref{eq:reverse_transition_kernel} with the posterior score. The posterior sampling process for the VP SDE is as follows:

\begin{equation}
    \begin{split}
        \mathbf{x}_{i}=&\frac{1}{\sqrt{\alpha_{i+1}}} \mathbf{x}_{i+1}+\frac{1-\alpha_{i+1}}{\sqrt{\alpha_{i+1}}} \mathbf{s}_{\theta}\left(\mathbf{x}_{i+1}, i+1\right) \\
        &+\sqrt{\frac{(1-\alpha_{i})(1-\bar{\alpha}_{i-1})}{1-\bar{\alpha}_{i}}} \mathbf{z} \\
        &+\frac{1-\alpha_{i+1}}{ \sqrt{\alpha_{i+1}}}\nabla_{\mathbf{x}_{i+1}}\log p(\mathbf{y}\vert\hat{\mathbf{x}}_{0}).\\
        \\
&\quad i = N-1, \dots, 0
    \end{split}
\end{equation}
The likelihood score can be scaled up by a factor of $\zeta>1$:
\begin{equation}
\begin{split}
        \nabla_{\mathbf{x}_{i}}\log \tilde{p}(\mathbf{x}_{i}\vert\mathbf{y}) &= \nabla_{\mathbf{x}_{i}}\log p(\mathbf{x}_{i}) + \zeta\cdot\nabla_{\mathbf{x}_{i}}\log p(\mathbf{y}\vert\mathbf{x}_{i} ) \\
        &= \nabla_{\mathbf{x}_{i}}\log p(\mathbf{x}_{i}) + \nabla_{\mathbf{x}_{i}}\log \frac{1}{Z}p(\mathbf{y}\vert\mathbf{x}_{i} )^{\zeta} \\
        &\approx \mathbf{s}_{\theta}(\mathbf{x}_{i},i) + \nabla_{\mathbf{x}_{i}}\log \frac{1}{Z}p(\mathbf{y}\vert\hat{\mathbf{x}}_{0})^{\zeta},
\end{split}
\end{equation}
where $Z$ is an arbitrary normalizing constant. Then the CKM can be drawn from a sharper distribution
\begin{equation}\label{eq:sharper_distribution}
    \tilde{p}(\mathbf{x}\vert \mathbf{y})\triangleq \frac{1}{Z}p(\mathbf{x})p(\mathbf{y}\vert\mathbf{x} )^{\zeta}
\end{equation}
which generates a CKM with higher fidelity and lower diversity \cite{dhariwal2021diffusion}. As discussed in Section~\ref{sec:Introduction}, in the CKM construction problem, both high fidelity and low diversity are desirable since the solution is expected to be unique and correct.

In practice, the sampling process is divided into two parts: \emph{prior regularization} and \emph{observation constraint}. For example, the sampling process with a scaled-up likelihood of the VP SDE can be divided into
\begin{equation} \label{eq:sampling_process}
\begin{split}
\mathbf{x}_{i} &=
\underbrace{%
  \begin{aligned}[t]
    &\frac{1}{\sqrt{\alpha_{i+1}}}\mathbf{x}_{i+1}
    + \frac{1-\alpha_{i+1}}{\sqrt{\alpha_{i+1}}}\,\mathbf{s}_{\theta}(\mathbf{x}_{i+1}, i+1) \\
    &\quad + \sqrt{\tfrac{(1-\alpha_{i})(1-\bar{\alpha}_{i-1})}{1-\bar{\alpha}_{i}}}\,\mathbf{z}
  \end{aligned}
}_{\text{prior regularization}} \\
&\quad
\underbrace{+\frac{1-\alpha_{i+1}}{\sqrt{\alpha_{i+1}}}\cdot\zeta\cdot\nabla_{\mathbf{x}_{i+1}}\log p(\mathbf{y}\vert\hat{\mathbf{x}}_{0})}
_{\text{observation constraint}}. 
\end{split}
\end{equation}
 The prior regularization steers the sample toward high-probability regions of the CKM prior distribution $p(\mathbf{x})$, so that the generated CKM respects the generic spatial-spectral regularities of the radio environment, as discussed in Section~\ref{section:cloud_side}. In contrast, the observation constraint increases the likelihood $p(\mathbf{y}\vert\mathbf{x})$ of the constructed CKM by minimizing the measurement residual $\| \mathbf{y}-\mathcal{A}(\hat{\mathbf{x}}_{0}) \|_2^2$, thereby enforcing that the constructed CKM agrees with the observation $\mathbf{y}$.

\subsubsection{Prior Regularization}

 Notice that the CKM construction is a noisy inverse problem, and the approximation in Eq.~\eqref{posterior_approximation} inevitably propagates the measurement noise $\mathbf{n}$ into the observation constraint, as shown by plugging the observation process \eqref{eq:observation_process} into the approximation in Eq.~\eqref{posterior_approximation}. 
\begin{equation}
    \nabla_{\mathbf{x}_{i}}\lVert \mathbf{y} - \mathcal{A}(\hat{\mathbf{x}}_{0}) \rVert_{2}^{2} = \nabla_{\mathbf{x}_{i}}\lVert \mathcal{A}(\mathbf{x}_{0}) - \mathcal{A}(\hat{\mathbf{x}}_{0}) + \mathbf{n} \rVert_{2}^{2}.
\end{equation}
By contrast, the prior regularization uses a model that produces a clean CKM prior score, thereby providing an inherent denoising effect. To counteract the noise, we adopt a hybrid sampling algorithm termed predictor-corrector (PC) sampling~\cite{song2021score} to implement the prior regularization, instead of the plain ancestral sampling. One iteration of PC sampling consists of one step of ancestral sampling as the predictor and several steps of Langevin MCMC \cite{parisi1981correlation} as the corrector.

\subsubsection{Observation Constraint}

Let $\mathbf{x}_{i}'$ denote the output of the prior regularization. The standard observation constraint for the VP SDE in Eq.~\eqref{eq:sampling_process} is as follows:

\begin{equation}
\begin{split}
        \mathbf{x}_{i} = \mathbf{x}_{i}'-\frac{1-\alpha_{i+1}}{ \sqrt{\alpha_{i+1}}}\cdot\zeta\cdot\frac{1}{2 \sigma^{2}}\nabla_{\mathbf{x}_{i+1}}\left\Vert\mathbf{y}-\mathcal{A}\left(\hat{\mathbf{x}}_{0}\right)\right\rVert_{2}^{2}.& \\
\end{split}
\end{equation}
Let $\mathbf{L}\triangleq\frac{1}{2 \sigma^{2}}\left\Vert\mathbf{y}-\mathcal{A}\left(\hat{\mathbf{x}}_{0}\right)\right\rVert_{2}^{2}$, it follows that
\begin{equation}
    \frac{1}{2 \sigma^{2}}\nabla_{\mathbf{x}_{i}}\left\Vert\mathbf{y}-\mathcal{A}\left(\hat{\mathbf{x}}_{0}\right)\right\rVert_{2}^{2} = \frac{1}{2 \sigma^{2}}\frac{\partial\mathbf{L}}{\partial\hat{\mathbf{x}}_{0}(\mathbf{x}_{i})}\cdot\frac{\partial\hat{\mathbf{x}}_{0}(\mathbf{x}_{i})}{\partial\mathbf{x}_{i}}.
\end{equation}
Notice that $\frac{\partial\hat{\mathbf{x}}_{0}(\mathbf{x}_{i})}{\partial\mathbf{x}_{i}}$ takes the form of 
\begin{equation}
        \frac{\partial\hat{\mathbf{x}}_{0}(\mathbf{x}_{i})}{\partial\mathbf{x}_{i}} =\frac{1}{\sqrt{\bar{\alpha}_{i}}}(\mathbf{I} + (1-\bar{\alpha}_{i})\frac{\partial\mathbf{s}_{\theta}(\mathbf{x}_{i},i)}{\partial\mathbf{x}_{i}}),
\end{equation}
which implies that when the measurement noise level $\sigma$ is too small or the schedule coefficient ($\frac{1}{\sqrt{\bar{\alpha}_{i}}}$) is too large, the gradient $\nabla_{\mathbf{x}_{i}}\mathbf{L}$ will explode, making the sampling process hard to converge. For numerical stability, the gradient $\nabla_{\mathbf{x}_{i}}\lVert \mathbf{y} - \mathcal{A}(\hat{\mathbf{x}}_{0}) \rVert_{2}^{2}$ is normalized by the norm $\lVert \mathbf{y} - \mathcal{A}(\hat{\mathbf{x}}_{0}) \rVert_{2}$, and the other coefficients are eliminated. This normalization makes the observation constraint robust to the measurement noise level $\sigma$.
 The sampling algorithm for the VP SDE is summarized in Algorithm \ref{Sampling VP}. For completeness, we derive and implement the training and inference algorithm for the VE SDE, and provide the implementation in our GitHub repository.

\begin{algorithm}[t]
  \caption{CKM construction at the edge}\label{Sampling VP}
  \begin{algorithmic}
    \Require forward operator $\mathcal{A}(\cdot)$, noise schedule $\{\alpha_{i},\bar{\alpha}_{i}\}_{i=1}^{N}$, observation $\mathbf{y}$, number of timesteps $N$, number of corrector steps $M$, Langevin MCMC step size $\{\epsilon_{i}\}_{i=1}^{N}$, observation constraint strength $\zeta$
    \State $\mathbf{x}_{N}\sim\mathcal{N}(\mathbf{0},\mathbf{I})$
    \For{$i = N,N-1,\dots,1$}
      \State $\hat{\mathbf{x}}_{0}= \frac{1}{\sqrt{\bar{\alpha}_{i}}}(\mathbf{x}_{i} + (1-\bar{\alpha}_{i})\mathbf{s}_{\theta}(\mathbf{x}_{i}, i))$\Comment{progressive estimate}
        \State $\mathbf{z}\sim \mathcal{N}(\mathbf{0},\mathbf{I})$
        \State $\mathbf{x}_{i}' = \frac{1}{\sqrt{\alpha_{i}}}\mathbf{x}_{i} + \frac{1-\alpha_{i}}{\sqrt{\alpha_{i}}}\mathbf{s}_{\theta}(\mathbf{x}_{i},i) + \sqrt{\frac{(1-\alpha_{i})(1-\bar{\alpha}_{i-1})}{1-\bar{\alpha}_{i}}}\mathbf{z}$
        \Comment{ancestral sampling, predictor}
      \For{$j = 1, 2, \dots,M$}
        \State $\mathbf{z}\sim \mathcal{N}(\mathbf{0},\mathbf{I})$
        \State $\mathbf{x}_{i}' \gets \mathbf{x}_{i}' + \epsilon_{i}\mathbf{s}_{\theta}(\mathbf{x}_{i},i) + \sqrt{2\epsilon_{i}}\mathbf{z}$
        \Comment{Langevin MCMC, corrector}
      \EndFor
      \State $\mathbf{x}_{i-1}\gets \mathbf{x}_{i}'-\zeta\frac{\nabla_{\mathbf{x}_{i}}\lVert \mathbf{y} - \mathcal{A}(\hat{\mathbf{x}}_{0}) \rVert_{2}^{2}}{\lVert \mathbf{y} - \mathcal{A}(\hat{\mathbf{x}}_{0}) \rVert_{2}}$
      \Comment{observation constraint}
    \EndFor
    \State \Return $\mathbf{x}_0$
  \end{algorithmic}
\end{algorithm}

\section{Comparative Study and Numerical Results}\label{sec:comparative_study}

\subsection{Dataset}
\begin{figure}[h]
    \centering
    \subfloat[visualization of the AoA map]{%
        \includegraphics[width=0.48\linewidth]{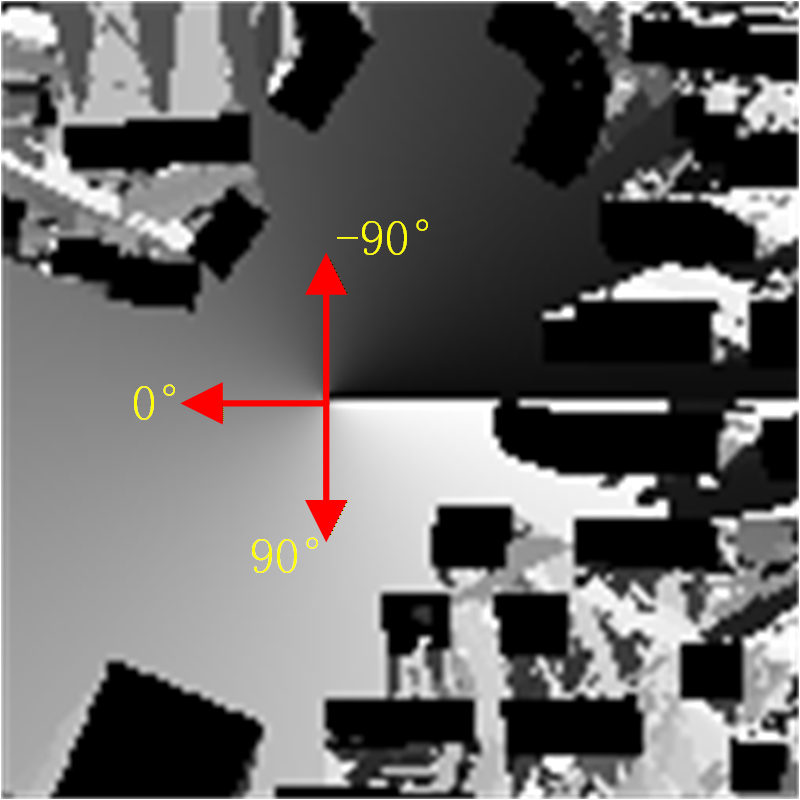}%
        \label{example_AoA}
    }
    \hfill
    \subfloat[visualization of the AoA-sine map]{%
        \includegraphics[width=0.48\linewidth]{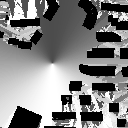}%
        \label{example_sin}
    }
    \caption{Comparison of the AoA map and the AoA-sine map.}
    \label{example_AoA_sin}
\end{figure}

We employ the CKMImageNet \cite{wu2025ckmimagenet} dataset to evaluate the proposed algorithm.  Specifically, the channel gain map (CGM) and the AoA map of the corresponding region are used as the training and testing data. To ensure that the model learns generalizable propagation patterns rather than memorizing specific maps, the training and testing sets are constructed from non-overlapping physical regions.
 For the CGM, the channel gain in $\left[-250, -50\right]$\,dB is linearly mapped to pixel values in $\left[0, 1\right]$. Any pixel corresponding to a building is assigned the minimum pixel value, indicating the absence of a received signal. For the AoA map, the original dataset stores the AoA of the strongest propagation path. In this work, however, only the sinusoidal value of the AoA is used for two reasons. First, a steering vector for a uniform linear array (ULA) with $N$ antenna elements, inter-element spacing $d$, and wavelength $\lambda$ can be written as
\begin{equation}
    \begin{split}
        \alpha(\theta)=\left[ 1,  e^{j\frac{2\pi}{\lambda}d\sin(\theta)},\dots, e^{j\frac{2\pi}{\lambda}d(N-1)\sin(\theta)}\right]^{T},
    \end{split}
\end{equation}
which depends only on the sinusoidal value of AoA $\sin(\theta)$ rather than the AoA $\theta$ itself. Second, representing the AoA in the pixel space requires mapping each angle to a pixel value. In the original dataset, the AoA $\theta$ is restricted to $\left[ -180^{\circ}, 180^{\circ}\right]$, and the region occupied by buildings is assigned a special angle value of $-200^{\circ}$ and is mapped to a pixel value of $0$. The physically meaningful angles $\left[ -180^{\circ}, 180^{\circ}\right]$ are mapped to the pixel values $\left[\frac{-180-(-200)}{180-(-200)}, 1\right]=\left[\frac{1}{19},1\right]$. The gap between $0$ and $\frac{1}{19}$ separates the building regions (no signal) from regions with a valid signal. As illustrated in Fig.~\ref{example_AoA}, the exact point where the black, gray, and white regions converge marks the BS. Due to the spatial periodicity of the AoA, $\theta=-180^{\circ}$ and $\theta=180^{\circ}$ correspond to the same physical direction, while their mapped pixel values abruptly change from $\frac{1}{19}$ to $1$. This representation is numerically correct, but semantically inconsistent and counterintuitive. To address this issue, we instead use the sine of the AoA. The value $\sin(\theta)\in \left[ -1, 1\right]$ is linearly mapped to the pixel value of $\left[0.3, 1\right]$, and pixels corresponding to buildings are assigned a pixel value of $0$. Again, the gap between $0$ and $0.3$ separates the building regions from the signal coverage. In this representation, the pixel value varies smoothly as the angle changes over one period, which yields a more interpretable pattern and is empirically easier for the model to learn, as shown in Fig.~\ref{example_sin}.

The channel metrics in the earlier radio map dataset~\cite{yapar2022dataset} are obtained by non-coherently summing the magnitudes of individual multipath components and therefore miss the fine-grained constructive and destructive interference that gives rise to fast fading (see Fig.~24 in the Sionna RT report~\cite{aoudia2025sionna}). In contrast, CKMImageNet retains individual multipath components and supports coherent summation, thereby preserving the small-scale fading statistics that are inevitably smoothed out in non-coherent maps and making it a more faithful benchmark. In this work, we use both the CGM and the AoA map of the same region to construct a two-channel training sample.

\subsection{Comparative study}

\begin{figure}[h]
    \centering
    \includegraphics[width=0.9\linewidth]{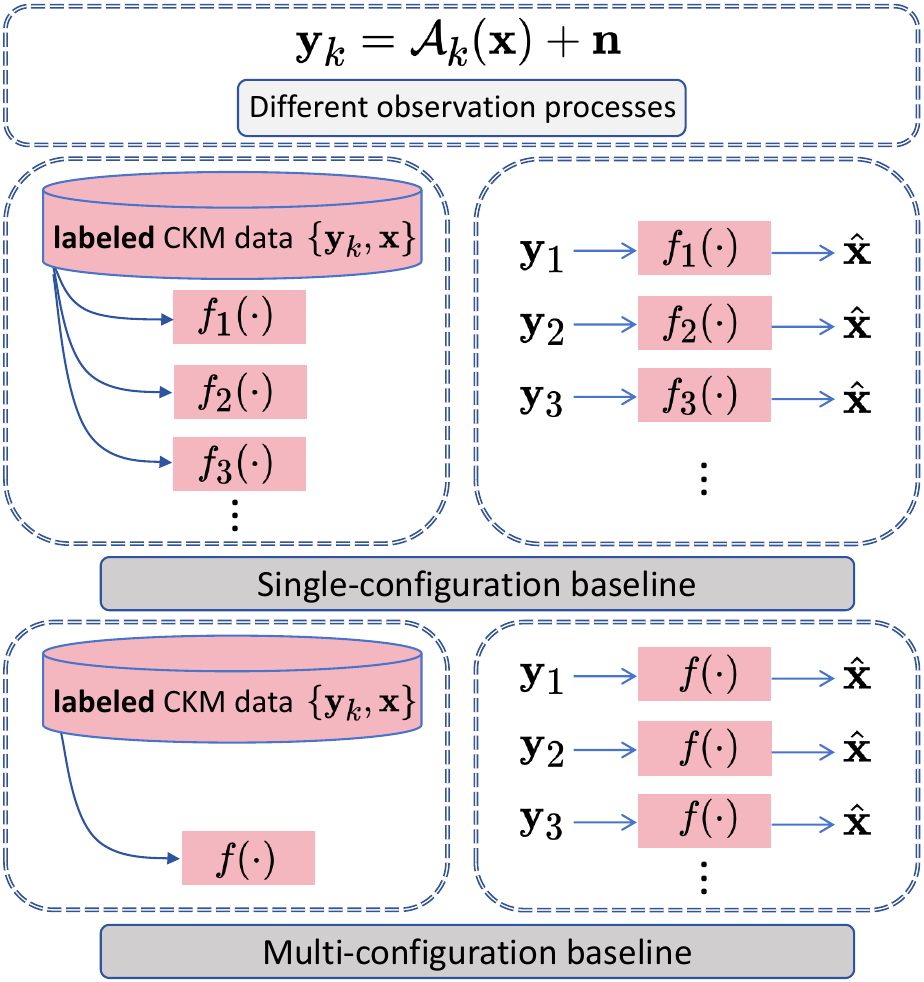}
    \caption{Setups of end-to-end baselines.}
    \label{fig:baseline_setup}
\end{figure}

We evaluate the proposed approach in terms of efficiency and accuracy. End-to-end schemes are chosen as baselines, and their setups are illustrated in Fig.~\ref{fig:baseline_setup}.

\subsubsection{Efficiency}
 
\begin{table*}[t]
\centering
\renewcommand{\arraystretch}{1.2}
\setlength{\tabcolsep}{5pt}
\caption{Efficiency and scalability comparison of different methods.}
\label{tab:efficiency}
\begin{tabular}{|c|c|c|c|}
\hline
\diagbox{Aspect}{Method} 
 & \makecell{\textbf{Proposed} \\ \textbf{foundation model}}
 & \makecell{Single-configuration \\ end-to-end model}
 & \makecell{Multi-configuration \\ end-to-end model} \\
\hline
\makecell{No. of large-scale training\\ runs for $k$ configurations}
 & $1$ 
 & $k$ 
 & $1$ (for current $k$ configurations) \\
\hline
\makecell{Additional cost \\ for a new configuration}
 & \textbf{0} 
 & New model and training 
 & Retraining / fine-tuning \\ 
\hline
\makecell{Generalization \\ to new configurations}
 & \textbf{Zero-shot}
 & No 
 & No \\
\hline
\makecell{Requirement  for labeled data \\ during training }
 & \textbf{No}
 & Yes 
 & Yes \\
\hline
\makecell{Performance \\ scaling w.r.t. $k$}
 & \textbf{Stable} (configuration-agnostic prior)
 & N/A (single-configuration) 
 & \makecell{Decreases as $k$ increases} \\
\hline

\multicolumn{4}{l}{\footnotesize Note: A \emph{configuration} refers to a specific setting of an inverse problem instance, including the task, hardware constraints, and noise level.}

\end{tabular}
\end{table*}

Table~\ref{tab:efficiency} compares the three paradigms in terms of training cost, data requirements, and scalability at a system level. In the proposed paradigm, the heavy training of the observation-agnostic CKM prior is performed once in the cloud with rich computing resources, using only unlabeled CKM data collected from multiple regions. The foundation model is then distributed to BSs equipped with edge servers, which only need to define the configuration-specific forward operators and run inference locally. When a new task or hardware constraint emerges, only the corresponding forward operator needs to be specified at the edge, without any additional heavy training or model synchronization in the cloud, so that the learned prior can be used in a zero-shot manner.

By contrast, the single-configuration end-to-end scheme couples the prior and likelihood within a black-box model for each configuration. For $k$ different configurations, the cloud has to maintain $k$ separate large models, each trained on its own labeled dataset. Any change in the sensing pipeline (e.g., a new type of degradation, a modified receiver front end, or a new noise level) brings about a new round of data collection, centralized training, and model deployment. This growth of training runs and model variants quickly becomes impractical when the construction configuration set expands.

The multi-configuration end-to-end scheme reduces the number of models to one, which trains a model on the union of all labeled datasets. However, it entangles all configurations within a single parameter set. Adding a new task or changing an existing degradation operator requires retraining or fine-tuning of the global backbone in the cloud, followed by the redeployment of the updated large model to all BSs. Moreover, such joint training suffers from negative transfer~\cite{standley2020tasks} across different configurations, and does not offer a principled way of knowledge sharing.

Overall, the foundation model paradigm is naturally aligned with a cloud--edge collaborative architecture: the cloud is responsible for training and periodically refreshing a task-agnostic CKM prior, while edge nodes perform lightweight inference by combining this shared prior with their local observations. This “train once in the cloud, adapt everywhere at the edge” workflow turns the CKM prior into a reusable knowledge base and enables the scalable and long-term evolution of tasks and devices. In contrast to the multi-configuration end-to-end scheme, the proposed method incorporates the shared prior with each configuration-specific observation through a likelihood term at the edge independently, so introducing new configurations does not lead to negative transfer.

\subsubsection{Accuracy}
\begin{figure*}[h]
  \centering
  \def\imgw{0.15\linewidth}

\begin{tabular}{@{}c@{\hspace{10pt}}c@{\hspace{10pt}}c@{\hspace{10pt}}c@{\hspace{10pt}}c@{}}
  \includegraphics[width=\imgw]{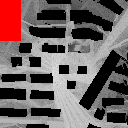} &
  \includegraphics[width=\imgw]{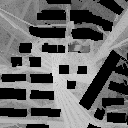} &
  \includegraphics[width=\imgw]{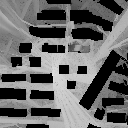} &
  \includegraphics[width=\imgw]{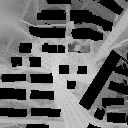} &
  \includegraphics[width=\imgw]{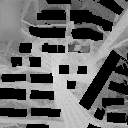} \\
  \\[-6pt]
  \includegraphics[width=\imgw]{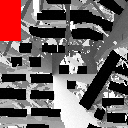} &
  \includegraphics[width=\imgw]{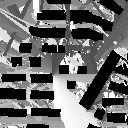} &
  \includegraphics[width=\imgw]{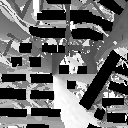} &
  \includegraphics[width=\imgw]{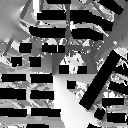} &
  \includegraphics[width=\imgw]{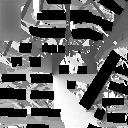} \\
  \\[-2pt]
  \scriptsize\parbox[c]{\imgw}{\centering\bfseries Observation} &
  \scriptsize\parbox[c]{\imgw}{\centering\bfseries Ground truth} &
  \scriptsize\parbox[c]{\imgw}{\centering\bfseries Proposed} &
  \scriptsize\parbox[c]{\imgw}{\centering\bfseries Single-config. Baseline} &
  \scriptsize\parbox[c]{\imgw}{\centering\bfseries Multi-config. Baseline}
\end{tabular}

  \caption{Visualization results for inpainting with box masks (IPbox). \textbf{Top:} gain. \textbf{Bottom:} AoA-sine.}
  \label{fig:vis_ipbox}
\end{figure*}

\begin{figure*}[h]
  \centering
  \def\imgw{0.15\linewidth}

\begin{tabular}{@{}c@{\hspace{10pt}}c@{\hspace{10pt}}c@{\hspace{10pt}}c@{\hspace{10pt}}c@{}}
  \includegraphics[width=\imgw]{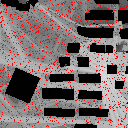} &
  \includegraphics[width=\imgw]{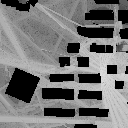} &
  \includegraphics[width=\imgw]{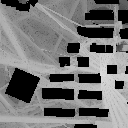} &
  \includegraphics[width=\imgw]{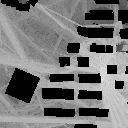} &
  \includegraphics[width=\imgw]{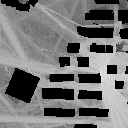} \\
  \\[-6pt]
  \includegraphics[width=\imgw]{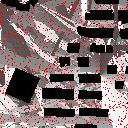} &
  \includegraphics[width=\imgw]{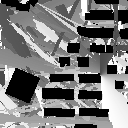} &
  \includegraphics[width=\imgw]{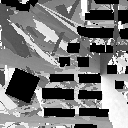} &
  \includegraphics[width=\imgw]{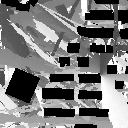} &
  \includegraphics[width=\imgw]{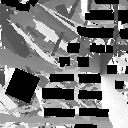} \\
  \\[-2pt]
  \scriptsize\parbox[c]{\imgw}{\centering\bfseries Observation} &
  \scriptsize\parbox[c]{\imgw}{\centering\bfseries Ground truth} &
  \scriptsize\parbox[c]{\imgw}{\centering\bfseries Proposed} &
  \scriptsize\parbox[c]{\imgw}{\centering\bfseries Single-config. Baseline} &
  \scriptsize\parbox[c]{\imgw}{\centering\bfseries Multi-config. Baseline}
\end{tabular}

  \caption{Visualization results for inpainting with random masks (IPrandom). \textbf{Top:} gain. \textbf{Bottom:} AoA-sine.}
  \label{fig:vis_iprandom}
\end{figure*}

\begin{figure*}[h]
  \centering
  \def\imgw{0.15\linewidth}

\begin{tabular}{@{}c@{\hspace{10pt}}c@{\hspace{10pt}}c@{\hspace{10pt}}c@{\hspace{10pt}}c@{}}
  \includegraphics[width=\imgw]{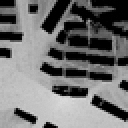} &
  \includegraphics[width=\imgw]{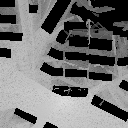} &
  \includegraphics[width=\imgw]{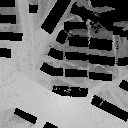} &
  \includegraphics[width=\imgw]{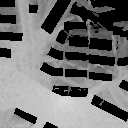} &
  \includegraphics[width=\imgw]{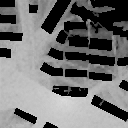} \\
  \\[-6pt]
  \includegraphics[width=\imgw]{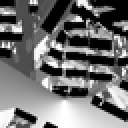} &
  \includegraphics[width=\imgw]{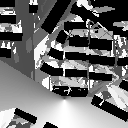} &
  \includegraphics[width=\imgw]{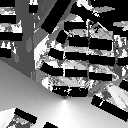} &
  \includegraphics[width=\imgw]{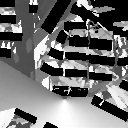} &
  \includegraphics[width=\imgw]{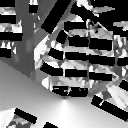} \\
  \\[-2pt]
  \scriptsize\parbox[c]{\imgw}{\centering\bfseries Observation} &
  \scriptsize\parbox[c]{\imgw}{\centering\bfseries Ground truth} &
  \scriptsize\parbox[c]{\imgw}{\centering\bfseries Proposed} &
  \scriptsize\parbox[c]{\imgw}{\centering\bfseries Single-config. Baseline} &
  \scriptsize\parbox[c]{\imgw}{\centering\bfseries Multi-config. Baseline}
\end{tabular}

  \caption{Visualization results for super-resolution (SR). \textbf{Top:} gain. \textbf{Bottom:} AoA-sine.}
  \label{fig:vis_sr}
\end{figure*}

\begin{figure*}[h]
  \centering
  \def\imgw{0.15\linewidth}

\begin{tabular}{@{}c@{\hspace{10pt}}c@{\hspace{10pt}}c@{\hspace{10pt}}c@{\hspace{10pt}}c@{}}
  \includegraphics[width=\imgw]{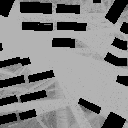} &
  \includegraphics[width=\imgw]{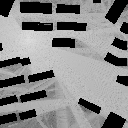} &
  \includegraphics[width=\imgw]{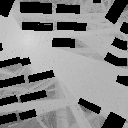} &
  \includegraphics[width=\imgw]{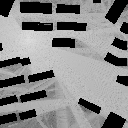} &
  \includegraphics[width=\imgw]{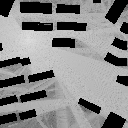} \\
  \\[-6pt]
  \includegraphics[width=\imgw]{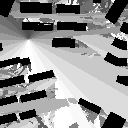} &
  \includegraphics[width=\imgw]{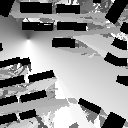} &
  \includegraphics[width=\imgw]{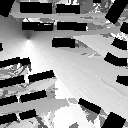} &
  \includegraphics[width=\imgw]{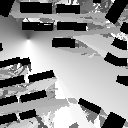} &
  \includegraphics[width=\imgw]{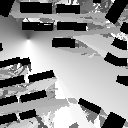} \\
  \\[-2pt]
  \scriptsize\parbox[c]{\imgw}{\centering\bfseries Observation} &
  \scriptsize\parbox[c]{\imgw}{\centering\bfseries Ground truth} &
  \scriptsize\parbox[c]{\imgw}{\centering\bfseries Proposed} &
  \scriptsize\parbox[c]{\imgw}{\centering\bfseries Single-config. Baseline} &
  \scriptsize\parbox[c]{\imgw}{\centering\bfseries Multi-config. Baseline}
\end{tabular}

  \caption{Visualization results for joint truncation and quantization recovery (JTQR). \textbf{Top:} gain. \textbf{Bottom:} AoA-sine.}
  \label{fig:vis_jtqr}
\end{figure*}

We evaluate the construction accuracy of the proposed method across four CKM construction configurations: (i) inpainting with a box mask (IPbox),
(ii) inpainting with a random mask (IPrandom),
(iii) super-resolution (SR), and
(iv) joint truncation and quantization recovery (JTQR). 
\begin{enumerate}
\item \textbf{Inpainting}

 Consider the case in~\ref{paragraph:masking}. Two types of masks are considered. For box masks, the masked region is defined as a rectangle with side lengths uniformly sampled from $\left[5,50\right]$ pixels. For random masks, the masking ratio is varied within $\left[\tfrac{5\times5}{128\times128}, \tfrac{50\times50}{128\times128}\right]\approx\left[0.001526,0.1526\right]$. The visualization results of the proposed method and baselines are shown in Fig.~\ref{fig:vis_ipbox} and Fig.~\ref{fig:vis_iprandom}.

\item \textbf{Super-resolution}

  Consider the case in~\ref{paragraph:downsampling}. The scale factor is set to $s=2$. The visualization results of the proposed method and baselines are shown in Fig.~\ref{fig:vis_sr}.

\item{\textbf{Joint truncation and quantization recovery}} 

Consider both cases in \ref{paragraph:truncation} and \ref{paragraph:quantization}, where the sensing device is affected by both dynamic range saturation on the gain measurement and finite angular resolution on the AoA measurement. Let $\mathbf{x}\in[0,1]^{2\times n}$ denote a sample with two data channels, where $x^{(g)}_{l}$ and $x^{(a)}_{l}$ are the pixel values of the gain map and the AoA-sine map at grid cell $l$, respectively. Pixels corresponding to buildings are encoded by special values in both layers and are not informative about the channel.

\paragraph*{Gain truncation}
On the gain layer, the truncation can be modeled as in Eq.~\eqref{eq:truncation_operator}
\begin{equation}
    \bigl[\mathcal{A}_{g}(\mathbf{x})\bigr]_l
    =
    \min\bigl(\max(x_l^{(g)}, a),\, b\bigr),
\end{equation}
where $a,b\in[0,1]$. We set $a=0.2$ and $b=0.7$ for the normalized pixel values. Using the linear mapping between the gain in dB and the pixel value,
\begin{equation}
    g_{\mathrm{dB}} = -250 + 200\,x^{(g)},
\end{equation}
the thresholds $a$ and $b$ correspond to $
    g_{\min}^{\mathrm{clip}}
    = -250 + 200 a
    = -210~\mathrm{dB}$ and $
    g_{\max}^{\mathrm{clip}}
    = -250 + 200 b
    = -110~\mathrm{dB}$,

i.e., channels weaker than $-210$\,dB or stronger than $-110$\,dB are saturated at the hardware output.

\paragraph*{Finite resolution quantization of the AoA}
On the AoA side, the sensing hardware can only resolve a finite number of angles. The full azimuthal range $[-180^{\circ},180^{\circ})$ is divided into $K$ equal sectors, and all angles within a sector are reported as the central angle of the sector. For example, when $K=12$, all angles within the sector $[0, 30^{\circ})$ are reported as $15^{\circ}$. Let $\theta_{l}\in[-180^{\circ},180^{\circ})$ denote the true AoA at grid cell $l$. Define
\begin{equation}
    \Delta = \frac{360^{\circ}}{K},\qquad
    \theta_{k} = -180^{\circ}+k\Delta,\quad k=0,\dots,K,
\end{equation}
as the sector boundaries, and
\begin{equation}
    \tilde{\theta}_{k}
    = -180^{\circ}+\Bigl(k+\tfrac{1}{2}\Bigr)\Delta,\quad k=0,\dots,K-1,
\end{equation}
as the corresponding reported angles. The angular quantizer implemented by the hardware is then
\begin{equation}
    Q_{\theta}(\theta)
    =
    \tilde{\theta}_{k},
    \quad
    \text{if } \theta \in [\theta_{k},\theta_{k+1}),\ k=0,\dots,K-1.
\end{equation}

We set $K=24$ in the experiments. According to the mapping rule of sinusoidal value to the pixel value, the forward operator on the AoA-sine layer in the pixel domain is therefore
\begin{equation}
    \bigl[\mathcal{A}_{a}(\mathbf{x})\bigr]_{l}
    =
    \begin{cases}
        0,\ \ \ \ \ \ \ \ \ \ \ \ \ \ \ \  \ \text{grid cell $l$ is building}  \\[2pt]
        0.35\sin \left(Q_\theta\left({\theta}_{l}\right)\right)+0.65, \ \ \ \text{otherwise}
    \end{cases},
\end{equation}
i.e., building locations are fixed to $0$, while non-building locations are uniformly quantized in the angular domain.

\paragraph*{Non-differentiability and the straight-through estimator}
Combining the gain layer and the AoA-sine layer, the joint truncation and quantization operator is
\begin{equation}
    \mathcal{A}(\mathbf{x})
    =
    \bigl(\mathcal{A}_{g}(\mathbf{x}),\,\mathcal{A}_{a}(\mathbf{x})\bigr).
\end{equation}
Due to the hard quantization in the AoA-sine branch, $\mathcal{A}(\cdot)$ is piecewise constant with respect to the AoA input, and its gradient is zero almost everywhere. As a result, it cannot be directly used within diffusion posterior sampling, which relies on backpropagating through the forward operator.
We then implement the AoA-sine branch using a straight-through estimator (STE)~\cite{bengio2013estimating}. Let
\begin{equation}
    v_{l}
    =
    0.35\sin(\theta_{l})+0.65
\end{equation}
denote the AoA-sine pixel value before quantization at grid cell $l$, and
\begin{equation}
    v^{\mathrm{hard}}_{l}
    =
    0.35\sin\!\bigl(Q_{\theta}(\theta_{l})\bigr)+0.65
\end{equation}
denote the pixel value after quantization. In the forward pass, we always use $v^{\mathrm{hard}}_{l}$, while in the backward pass, we approximate the gradient by treating the quantizer as the identity mapping. Let $\ell$ denote a scalar loss function used in diffusion posterior sampling. We then approximate
\begin{equation}
    \frac{\partial \ell}{\partial v_{l}}
    \approx
    \frac{\partial \ell}{\partial v^{\mathrm{hard}}_{l}}.
\end{equation}
To do so, a surrogate variable is introduced
\begin{equation}
    v^{\mathrm{STE}}_{l}
    =
    v_{l}
    +
    \bigl(v^{\mathrm{hard}}_{l} - v_{l}\bigr)_{\mathrm{stop\mbox{-}grad}},
\end{equation}
so that $v^{\mathrm{STE}}_{l} = v^{\mathrm{hard}}_{l}$ in the forward pass, but
$\partial v^{\mathrm{STE}}_{l} / \partial v_{l} = 1$ in the backward pass.  

The visualization results of the proposed method and baselines are shown in Fig.~\ref{fig:vis_jtqr}.
\end{enumerate}

For a fair comparison, the end-to-end baselines are implemented using the identical U-Net backbone of the foundation score model, with only minimal modifications to match the input and output formats, thus precluding the impact of model parameter count (model capacity) on the comparison. In the single-configuration end-to-end setting, four separate models are trained using the labeled dataset of each configuration. For the multi-configuration end-to-end baseline, a single model is trained on the union of the labeled datasets from the four configurations. In contrast, the proposed approach trains a single foundation model using an unlabeled CKM dataset.

Accuracy is evaluated in terms of the RMSE of the gain map (in dB) and the AoA-sine map, summarized in Tables~\ref{tab:performance_gain} and~\ref{tab:performance_sin}, respectively. The accuracy of the \emph{gain map} is obtained by scaling the RMSE of the gain in the pixel domain by $(-50)-(-250)=200$. The accuracy of the \emph{AoA-sine map} is assessed using the RMSE of $\sin(\theta)$, which quantifies the construction error of the directional component along the vertical axis. Since $\sin(\theta)\in[-1,1]$ is linearly mapped to the pixel range $[0.3,1]$, the RMSE in the $\sin(\theta)$ domain is obtained by multiplying the pixel-domain RMSE by $\frac{1-(-1)}{1-0.3} = \frac{20}{7}$. The noise scale is set to $\sigma=0.01$ in the pixel domain. As discussed in Section~\ref{sec:Introduction}, unlike natural image restoration, where diversity may exist for the same observation and perceptual metrics (e.g., SSIM or FID) are often preferred, CKM construction is a physics-grounded inference problem with a unique underlying ground truth. Therefore, we adopt a correctness-oriented metric (RMSE) to quantify construction accuracy.

\begin{table}[h]
\centering
\renewcommand{\arraystretch}{1.3} 
\setlength{\tabcolsep}{8pt}       
\caption{Accuracy comparison of gain: RMSE (dB).}
\label{tab:performance_gain}
\begin{tabular}{|c|c|c|c|c|}
\hline
\diagbox{Method}{Task} & IPbox & IPrandom & SR & JTQR \\
\hline
\makecell{\textbf{Proposed} \\\textbf{foundation model}} &10.082&8.772&7.921&3.483\\
\hline
\makecell{Single-configuration \\ end-to-end model} & 9.797&10.029&5.44&3.189\\
\hline
\makecell{Multi-configuration \\ end-to-end model} & 14.868 &13.201&6.176&3.550\\
\hline
\end{tabular}
\end{table}

\begin{table}[h]
\centering
\renewcommand{\arraystretch}{1.3} 
\setlength{\tabcolsep}{8pt}       
\caption{Accuracy comparison of AoA-sine: RMSE.}
\label{tab:performance_sin}
\begin{tabular}{|c|c|c|c|c|}
\hline
\diagbox{Method}{Task} & IPbox & IPrandom & SR & JTQR \\
\hline
\makecell{\textbf{Proposed} \\ \textbf{foundation model}}&0.378&0.381&0.208&0.052\\
\hline
\makecell{Single-configuration \\ end-to-end model} &0.401&0.370&0.158&0.050\\
\hline
\makecell{Multi-configuration \\ end-to-end model} &0.519&0.413&0.180&0.053\\
\hline
\end{tabular}
\end{table}

Overall, the single-configuration end-to-end model enjoys a slight advantage on most metrics. This is expected: under the MSE criterion, the optimal estimator is $\hat{\mathbf{x}}(\mathbf{y}) = \mathbb{E}\left[\mathbf{x}\vert\mathbf{y}\right]$, which a supervised model trained with MSE approximates for each fixed configuration~\cite{bishop2006prml}, as discussed in~\ref{para:MMSE_construction}. The proposed foundation model attains competitive performance across the four configurations: it achieves the best RMSE on the IPbox for the AoA-sine map and on the IPrandom for the gain map, while remaining close to the single-configuration baseline on the remaining configurations. It outperforms the multi-configuration end-to-end model on almost all metrics, which shows again that training a single supervised model on the
union of data from heterogeneous configurations induces negative transfer~\cite{standley2020tasks},
making it difficult for one shared set of parameters to fit the
data of all configurations equally well, though it still relies on a fully labeled dataset from all configurations. Despite not being explicitly optimized for MSE, the proposed prior–likelihood decoupled paradigm can deliver strong construction accuracy while retaining advantages in data requirements and scalability.

\subsection{\texorpdfstring{Sensitivity to the Observation Constraint Strength $\zeta$}{Sensitivity to the Observation Constraint Strength zeta}}

This subsection investigates how the observation constraint strength $\zeta$ in the posterior inference algorithm affects the CKM construction performance for different configurations. For each construction configuration, we evaluate the proposed algorithm under a set of candidate values of $\zeta$ and report the corresponding RMSE. The value of $\zeta$ that yields the lowest RMSE is adopted as the default setting in the other experiments.
\begin{figure}[h]
    \centering
    \includegraphics[width=0.95\linewidth]{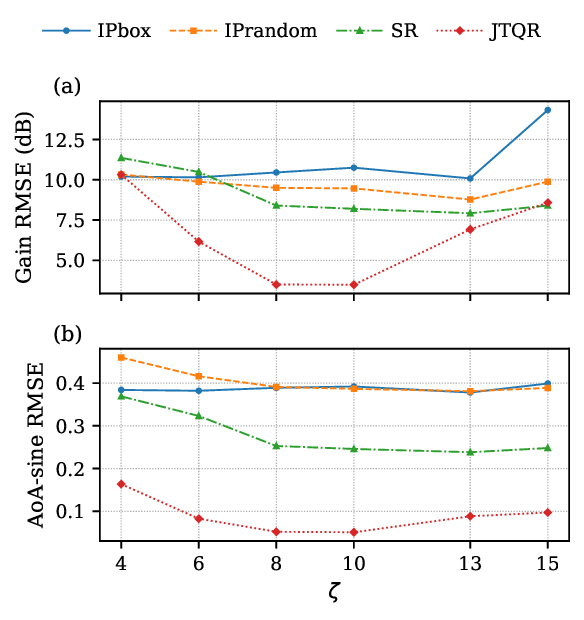}
    \caption{Sensitivity of CKM construction accuracy to the observation constraint strength $\zeta$.}
    \label{fig:zeta_sensitivity}
\end{figure}
Fig.~\ref{fig:zeta_sensitivity} illustrates the sensitivity of the construction accuracy to $\zeta$. Three configurations (IPbox, IPrandom, and SR)
all reach their minimum RMSE at $\zeta=13$, whereas the JTQR achieves the best performance at
$\zeta=10$.
Overall, the sweet spot is attained at relatively large values
of $\zeta$. These results indicate that the observation constraint needs to be weighted much
more heavily than the learned prior in the posterior sampling.
As shown in Eq.~\eqref{eq:sharper_distribution}, the effective posterior behaves like a sharp distribution whose mass is
tightly clustered around a single point.
This behavior is consistent with the inherent property of the considered inverse
problems discussed in Section~\ref{sec:Introduction}: under a certain measurement model, the underlying CKM is
uniquely determined by the observations, so that a highly
peaked posterior and limited diversity are desirable.

On the other hand, the curves also show that when $\zeta$ becomes too large, the RMSE starts to increase again.
This suggests that excessively emphasizing the observation constraint  makes
the sampler overfit to the measurement noise, and also amplifies the approximation error of the likelihood score.
Therefore, a finite but relatively large value of $\zeta$ reaches a balance
between enforcing observation consistency and exploiting the learned prior, and
the range $\zeta\in[10,13]$ provides a robust operating region shared by
all configurations. In contrast, for natural image inverse problems, a smaller observation constraint strength is typically used (e.g., $\zeta \in\left[0.3, 1\right]$) \cite{chung2023diffusion} to optimize perceptual metrics.

Another observation from Fig.~\ref{fig:zeta_sensitivity} is that the IPbox is
relatively insensitive to the choice of $\zeta$. This behavior is
consistent with the ill-posedness of the underlying inverse problems. For the IPbox, the forward operator $\mathcal{A}(\cdot)$ is highly underdetermined, and the observations provide limited information, thus the
prior regularization plays a dominant role in determining the construction.
Consequently, adjusting the weight of the observation constraint within a
moderate range leads to minor changes in the RMSE. In contrast, the other configurations provide
more informative observations, and the posterior mass responds more
strongly to changes in $\zeta$, leading to a sharper optimum in terms of
RMSE.

\section{Conclusion}\label{sec:conclusion}

In this work, we present a cloud--edge collaborative framework for CKM construction that is explicitly designed for system-level scalability, thereby meeting the requirements of AI--RAN services. By factorizing the posterior into a configuration-agnostic CKM prior and a configuration-specific likelihood, we showed that the CKM prior can be learned once in the cloud as a foundation prior model using an unlabeled CKM dataset, and then disseminated to many BSs/edge nodes as a shared knowledge base. At the edge, CKM construction is performed via MAP-oriented posterior inference, which fuses the CKM prior with local observations in a plug-and-play manner. 

Extensive and controlled evaluations over both linear (inpainting, super-resolution) and nonlinear (joint truncation and quantization recovery) configurations verify that the proposed paradigm achieves competitive RMSE performance while offering clear practical benefits: (i) one-time large-scale training instead of per-configuration training; (ii) no dependence on labeled data; (iii) robustness against the negative transfer effect; and (iv) zero-shot adaptation to new configurations. Sensitivity analysis further indicates that appropriately weighting the observation constraint is important for correctness-oriented CKM construction, consistent with the fact that CKM inverse problems favor high-fidelity, low-diversity solutions. Overall, our work offers a principled and practically deployable step toward scalable and efficient CKM construction in future AI--native, cloud--edge collaborative 6G architectures.

\section*{Acknowledgment}
Computing resources were provided by the \href{https://nic.seu.edu.cn/}{Network and Information Center of Southeast University} and the \href{https://hiascend.com}{Huawei Ascend AI Computing Platform}.

\ifCLASSOPTIONcaptionsoff
  \newpage
\fi

\bibliographystyle{IEEEtran}
\bibliography{reference.bib}
\end{document}